\documentclass[conference]{IEEEtran}
\IEEEoverridecommandlockouts
\usepackage{booktabs}     
\usepackage{multirow}     
\usepackage{tabularx}     
\usepackage{pdflscape}    
\usepackage{array}        
\usepackage{caption}      
\usepackage{cite}         
\usepackage{amsmath,amssymb,amsfonts} 
\usepackage{algorithmic}  
\usepackage{graphicx}     
\usepackage{textcomp}     
\usepackage{xcolor}       
\usepackage{adjustbox}    
\usepackage{algorithm}
\usepackage{algorithmic}

\newcolumntype{Y}{>{\centering\arraybackslash}X}

\def\BibTeX{{\rm B\kern-.05em{\sc i\kern-.025em b}\kern-.08em
    T\kern-.1667em\lower.7ex\hbox{E}\kern-.125emX}}

\def\BibTeX{{\rm B\kern-.05em{\sc i\kern-.025em b}\kern-.08em
    T\kern-.1667em\lower.7ex\hbox{E}\kern-.125emX}}
\begin{document}
\vspace{-4mm}
\title{DD-JSCC: Dynamic Deep Joint Source-Channel Coding for Semantic Communications\\
\thanks{}
}


\vspace{-5mm}
\author{
    \IEEEauthorblockN{
        Avi Deb Raha\textsuperscript{1},
        Apurba Adhikary\textsuperscript{1},
        Mrityunjoy Gain\textsuperscript{2},
        Yumin Park\textsuperscript{1},
        Walid Saad\textsuperscript{3},
        and Choong Seon Hong\textsuperscript{1}\textsuperscript{*}
    }
    \IEEEauthorblockA{
        \textsuperscript{1}Department of Computer Science and Engineering, Kyung Hee University, Yongin-si 17104, Republic of Korea\\
        \textsuperscript{2}Department of Artificial Intelligence, Kyung Hee University, Yongin-si 17104, Republic of Korea\\
        \textsuperscript{3}Department of Electrical and Computer Engineering, Virginia Tech, Arlington, VA 22203, USA\\
        E-mail: \{avi, apurba, gain, yumin0906, cshong\}@khu.ac.kr, walids@vt.edu
    }
    \vspace{-10mm}
\thanks{This work has been accepted by the 2025 IEEE International Conference on Communications (IEEE ICC 2025), \textcopyright~2025 IEEE.
Copyright may be transferred without notice, after which this version may no longer be accessible. Personal use of this material is permitted.
Permission from IEEE must be obtained for all other uses, including reprinting/republishing for advertising or promotional purposes, creating new collective works, resale or redistribution to servers or lists, or reuse of any copyrighted component of this work in other works.
*Dr. CS Hong is the corresponding author.}
}


\maketitle
\vspace{-10mm}
\begin{abstract}
Deep Joint Source-Channel Coding (Deep-JSCC) has emerged as a promising semantic communication approach for wireless image transmission by jointly optimizing source and channel coding using deep learning techniques. However, traditional Deep-JSCC architectures employ fixed encoder-decoder structures, limiting their adaptability to varying device capabilities, real-time performance optimization, power constraints and channel conditions. To address these limitations, we propose DD-JSCC: Dynamic Deep Joint Source-Channel Coding for Semantic Communications, a novel encoder-decoder architecture designed for semantic communication systems. Unlike traditional Deep-JSCC models, DD-JSCC is flexible for dynamically adjusting its layer structures in real-time based on transmitter and receiver capabilities, power constraints, compression ratios, and current channel conditions. This adaptability is achieved through a hierarchical layer activation mechanism combined with implicit regularization via sequential randomized training, effectively reducing combinatorial complexity, preventing overfitting, and ensuring consistent feature representations across varying configurations. Simulation results demonstrate that DD-JSCC enhances the performance of image reconstruction in semantic communications, achieving up to 2 dB improvement in Peak Signal-to-Noise Ratio (PSNR) over fixed Deep-JSCC architectures, while reducing training costs by over 40\%. The proposed unified framework eliminates the need for multiple specialized models, significantly reducing training complexity and deployment overhead.
\end{abstract}


\section{Introduction}
The rapid proliferation of connected devices and the emergence of data-intensive applications such as autonomous vehicles \cite{Avi_SemCom_ICOIN}, real-time financial trading, online gaming, and advanced telecommunications have dramatically increased the demands on wireless communication networks\cite{qiao2025deepseek}. These applications require communication systems that are not only high in speed and reliability but also highly adaptable to dynamic and heterogeneous operational environments. Traditional communication paradigms, which typically separate source coding and channel coding processes, are increasingly inadequate in meeting these sophisticated requirements. In conventional wireless systems, source coding techniques (e.g., JPEG, BPG, JPEG2000) compress data, while channel coding methods (e.g., LDPC, Polar, Turbo codes) ensure transmission reliability~\cite{PADC}. However, this layered approach optimizes source and channel coding independently, leading to suboptimal communication capacity and the "cliff-effect," where data quality sharply degrades below certain SNR thresholds~\cite{PADC}.

To address these limitations, Joint Source-Channel Coding (JSCC) has emerged as a paradigm that integrates source and channel coding into a unified framework to optimize the end-to-end communication process~\cite{jscc}. Recent advancements in deep learning have further propelled JSCC methodologies, leading to Deep Joint Source-Channel Coding (Deep-JSCC) models that leverage deep neural networks, particularly autoencoders, to enhance both data compression efficiency and transmission reliability~\cite{djscc}. Deep-JSCC models have demonstrated significant improvements over traditional methods in various wireless data transmission tasks, including image~\cite{djscc, PADC}, text~\cite{tdjscc}, and speech~\cite{9450827} transmission. These models effectively mitigate the cliff-effect by maintaining robust performance across a range of SNR conditions and compression ratios, thereby achieving closer to system-level optimality.

Despite these advancements, existing Deep-JSCC models predominantly rely on fixed encoder-decoder architectures that are not designed to adapt dynamically to changing device capabilities, power constraints, or varying channel conditions. In real-world wireless environments, especially in mobile and power-constrained devices, communication systems must contend with varying computational resources and fluctuating power levels. A fixed architecture poses significant inefficiencies in such scenarios. For instance, under high SNR conditions or when devices have limited computational resources, a lightweight model suffices, conserving energy and computational power. Conversely, in low SNR environments or when higher fidelity is required, a deeper model may be necessary to ensure reliable communication. Maintaining multiple fixed architectures to accommodate these varying conditions is impractical due to the substantial memory and computational overhead, particularly in heterogeneous networks where devices have different capabilities.

In this paper, we address these challenges by proposing a Dynamic Deep Joint Source-Channel Coding (DD-JSCC) architecture for semantic communication systems that can adapt its layer structure in real-time based on transmitter and receiver capabilities, power constraints, compression ratios, and current channel conditions. Unlike traditional approaches that require multiple specialized models for different operational scenarios~\cite{jiang2024semantic}, or methods that rely on dynamic partitioning \cite{ICC_bal} of fixed architectures, which remain effectively static by overburdening one device when both the sender and receiver have constrained computational resources, leading to performance bottlenecks and excessively large transmitted feature sizes, our DD-JSCC architecture employs a flexible layer configuration mechanism. This enables the encoder and decoder to dynamically activate or deactivate layers, ensuring balanced computational loads and optimized data compression. Consequently, a single model can efficiently operate across a wide spectrum of conditions, enhancing both data compression and transmission reliability without the need for maintaining multiple specialized models. The main contributions of this paper are summarized as follows:

\begin{enumerate}
    \item We propose a novel encoder-decoder architecture for semantic communication that is fexible for dynamically adjusting its layer structures in real-time based on channel SNR, compression ratio (CR), device capabilities, and power constraints. This adaptation ensures optimal resource utilization and maintains high performance across diverse operational conditions. To the best of our knowledge, this work is the first to introduce a fully dynamic architecture that enables semantic communication systems to optimize both transmitter and receiver configurations concurrently.

    \item We propose a novel training strategy that employs sequential randomized layer selection with hierarchical constraints. This method acts as an implicit regularizer, enhancing model generalization and robustness by preventing overfitting and promoting the learning of efficient feature representations across varying network configurations.

    \item Simulation results demonstrate that the proposed DD-JSCC architecture significantly enhances the performance of image reconstruction in semantic communications. Specifically, DD-JSCC achieves up to {2\,dB} improvement in Peak Signal-to-Noise Ratio (PSNR) over fixed Deep-JSCC architectures for image reconstruction tasks. Additionally, the DD-JSCC reduces training costs by over {40\%}, highlighting its efficiency and scalability for real-world deployments.
\end{enumerate}
\section{System Model and Problem Formulation} \label{smpf}
\vspace{-2mm}
In this section, we present a comprehensive framework for the proposed DD-JSCC architecture designed for semantic communication in wireless image transmission systems. 
\subsection{System Model}
\begin{figure}[t]
\centering
\includegraphics[width=9cm, height=7cm]{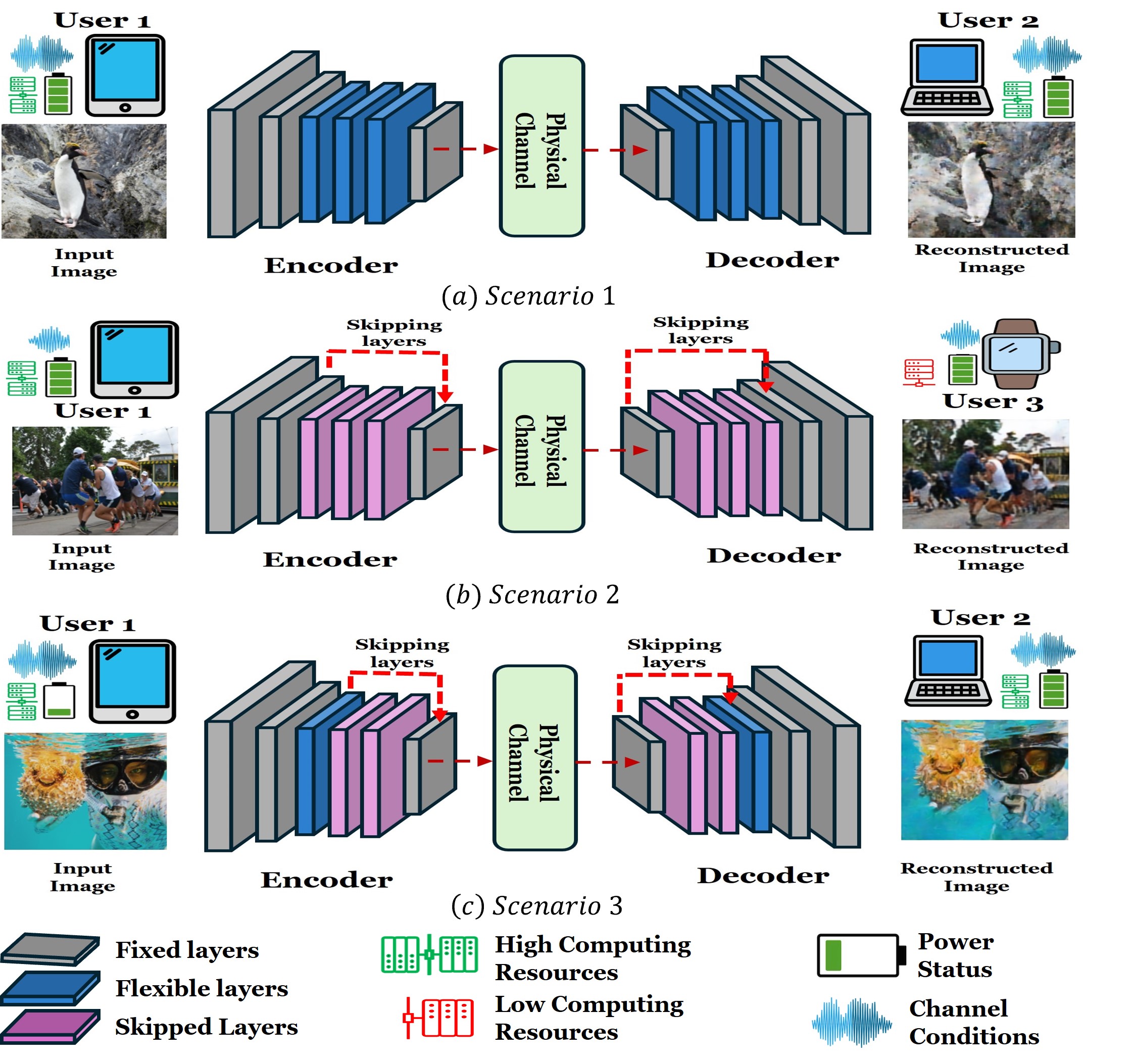}
\vspace{-2mm}
\caption{System Model for Dynamic Deep Joint Source-Channel
Coding for Semantic Communications.}
\vspace{-4mm}
\label{system_model}
\vspace{-2mm}
\end{figure}

The system model comprises three main components: the transmitter, the physical channel, and the receiver. The transmitter includes a semantic encoder, while the receiver is equipped with a decoder. The DD-JSCC architecture aims to reconstruct the original image $\mathbf{x} \in \mathbb{R}^{N}$ at the receiver with minimal distortion, dynamically adapting to varying channel conditions, device computational capabilities and power constraints as shown in fig. \ref{system_model}.
\subsubsection{Transmitter}
The transmitter consists of a semantic encoder $E_{\phi}(\cdot)$, which transforms the input image $\mathbf{x} \in \mathbb{R}^{N}$ into a lower-dimensional complex-valued semantic code $\mathbf{z} \in \mathbb{C}^{K}$:
\vspace{-2mm}
\begin{equation}
\mathbf{z} = E_{\phi}(\mathbf{x}; \gamma, R, \mathbf{L}_e),
\label{eq:encoder}
\vspace{-2mm}
\end{equation}
where, $\gamma$ denotes the Signal-to-Noise Ratio (SNR), which influences the encoding strategy based on current channel conditions. $R = \frac{K}{N}$ represents the desired Compression Ratio (CR). $\mathbf{L}_e = [l_1^e, l_2^e, \dots, l_L^e]$ is the encoder's active layer configuration vector, where each $l_i^e \in \{0,1\}$ indicates whether the $i^{th}$ layer is active ($1$) or inactive ($0$).
\vspace{-1mm}
\subsubsection{Power Normalization}
To comply with transmission power constraints, the semantic code $\mathbf{z}$ is normalized before transmission. The power-normalized semantic code $\mathbf{s} \in \mathbb{C}^{K}$ is given by:
\vspace{-2mm}
\begin{equation}
\mathbf{s} = \sqrt{\frac{K P_{\text{max}}}{\|\mathbf{z}\|^2}} \mathbf{z},
\label{eq:power_normalization}
\vspace{-2mm}
\end{equation}
where $\|\mathbf{z}\|^2 = \sum_{k=1}^{K} |z_k|^2$ and $P_{\text{max}}$ is the maximum allowed average power per symbol. This normalization ensures that the average power constraint is satisfied:
\begin{equation}
\vspace{-2mm}
\frac{1}{K} \mathbb{E}\left[ \|\mathbf{s}\|^2 \right] \leq P_{\text{max}}.
\label{eq:power_constraint}
\end{equation}

\begin{figure*}[t]
\centering
\includegraphics[width=15cm]{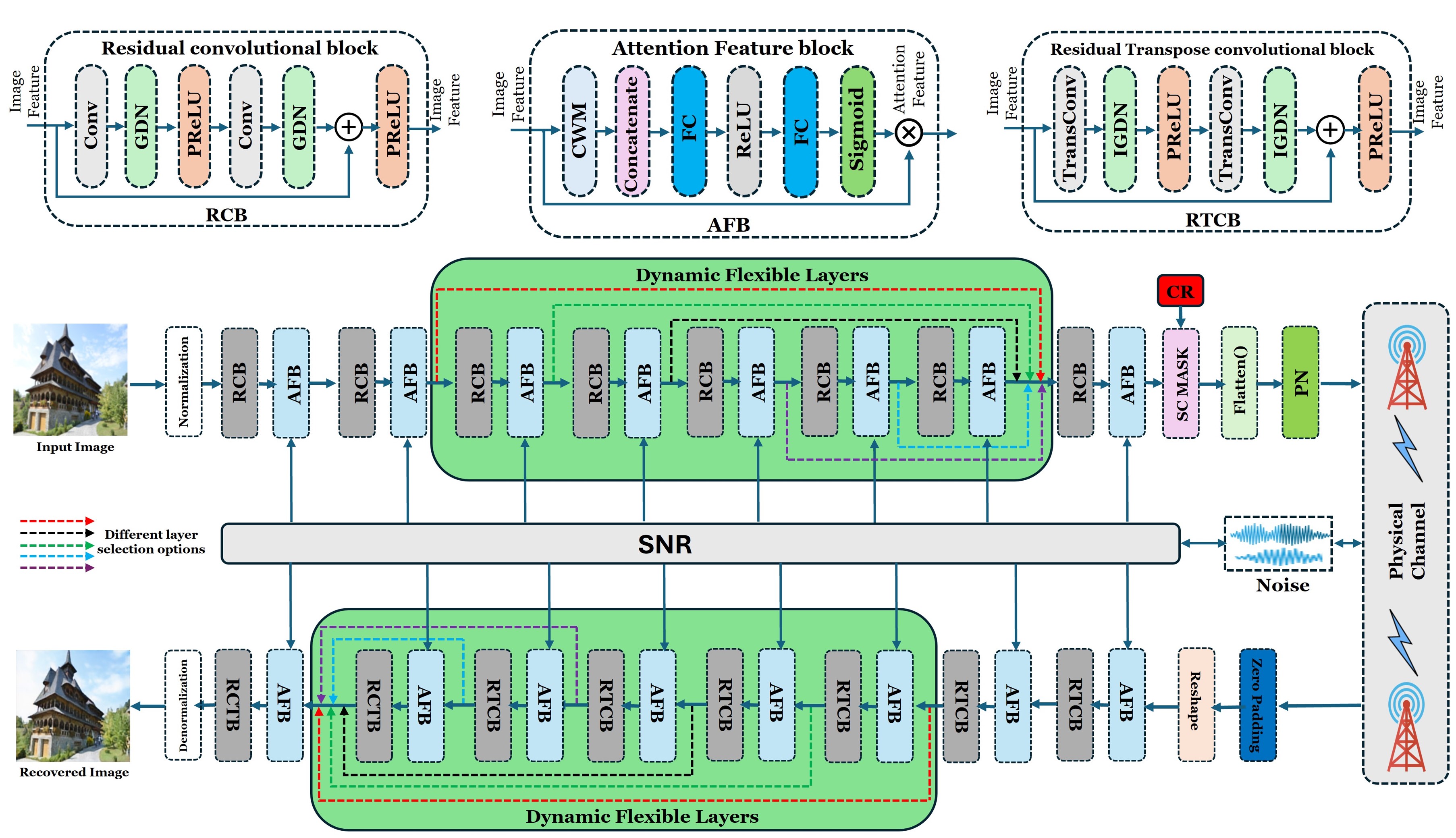}
\vspace{-2mm}
\caption{Overview of the proposed DD-JSCC architecture. The flexible DD-JSCC framework can be adapted to various network architectures. In this study, we demonstrate its application using the DeepJSCC-V architecture.}
\label{system_model_overview}
\vspace{-4mm}
\end{figure*}
\subsubsection{Physical Channel}
The normalized semantic code $\mathbf{s}$ is transmitted over a wireless channel modeled as:
\vspace{-2mm}
\begin{equation}
\tilde{\mathbf{z}} = h \mathbf{s} + \mathbf{n},
\label{eq:channel_model}
\vspace{-2mm}
\end{equation}
where $h \in \mathbb{C}$ represents the complex channel gain coefficient, capturing the effects of fading and path loss. $\mathbf{n} \sim \mathcal{CN}(\mathbf{0}, \sigma^2 \mathbf{I}_K)$ denotes the additive white Gaussian noise (AWGN) vector, with $\sigma^2$ being the noise variance.
For scenarios where the channel gain is normalized (i.e., $h = 1$), the channel model simplifies to:
\vspace{-2mm}
\begin{equation}
\tilde{\mathbf{z}} = \mathbf{s} + \mathbf{n}.
\label{eq:channel_model_simplified}
\vspace{-2mm}
\end{equation}
The channel SNR $\gamma$ is defined as:
\vspace{-1mm}
\begin{equation}
\gamma = \frac{\mathbb{E}\left[ |h|^2 \|\mathbf{s}\|^2 \right]}{\mathbb{E}\left[ \|\mathbf{n}\|^2 \right]} = \frac{|h|^2 P_{\text{max}}}{\sigma^2},
\label{eq:snr}
\vspace{-1mm}
\end{equation}
and is often expressed in decibels (dB) as:
\vspace{-1mm}
\begin{equation}
\gamma_{\text{dB}} = 10 \log_{10}(\gamma).
\vspace{-1mm}
\label{eq:snr_db}
\end{equation}
\subsubsection{Receiver}
The receiver comprises a decoder $D_{\theta}(\cdot)$, which reconstructs the image from the received signal $\tilde{\mathbf{z}}$:
\vspace{-1mm}
\begin{equation}
\hat{\mathbf{x}} = D_{\theta}(\tilde{\mathbf{z}}; \gamma, R, \mathbf{L}_d),
\label{eq:decoder}
\vspace{-2mm}
\end{equation}
where $\mathbf{L}_d = [l_1^d, l_2^d, \dots, l_L^d]$ is the decoder's active layer configuration vector, where each $l_i^d \in \{0,1\}$ indicates whether the $i^{th}$ layer is active ($1$) or inactive ($0$).
\vspace{-2mm}
\subsection{Loss Function}
\vspace{-2mm}
To train the DD-JSCC architecture, we minimize the reconstruction error between the original image $\mathbf{x}$ and the reconstructed image $\hat{\mathbf{x}}$. This error is quantified using the Mean Squared Error (MSE):
\vspace{-2mm}
\begin{equation}
\mathcal{L}(\mathbf{x}, \hat{\mathbf{x}}) = \frac{1}{N} \|\mathbf{x} - \hat{\mathbf{x}}\|^2 = \frac{1}{N} \sum_{i=1}^{N} (x_i - \hat{x}_i)^2.
\label{eq:loss_function}
\vspace{-2mm}
\end{equation}
\subsection{Problem Formulation}
\vspace{-2mm}
The encoder and decoder parameters $(\phi, \theta)$ are jointly optimized to minimize the expected reconstruction error across the distributions of the input images $\mathbf{x}$, channel SNRs $\gamma$, compression ratios $R$, and layer configurations $\mathbf{L}_e$ and $\mathbf{L}_d$. The optimization problem can be formulated as:
\begin{equation}
\min_{\phi, \theta} \ \mathbb{E}_{\mathbf{x}, \gamma, R, \mathbf{L}_e, \mathbf{L}_d} \left[ \mathcal{L}\left( \mathbf{x}, D_{\theta} \left( h\mathbf{s} + \mathbf{n}; \gamma, R, \mathbf{L}_d \right) \right) \right],
\label{eq:optimization_objective}
\end{equation}
where $\mathcal{L}(\cdot)$ denotes the reconstruction loss between the original image $\mathbf{x}$ and the reconstructed image $\hat{\mathbf{x}}$. This formulation aims to jointly train the encoder and decoder to minimize the average reconstruction loss over varying images, channel conditions, compression ratios, and layer configurations.

The dynamic activation of layers in both the encoder and decoder introduces substantial complexities to the training and optimization process. Specifically, the network parameters must be optimized across an exponentially large number of possible layer configurations due to the combinatorial nature of layer activations. This results in a highly non-convex and intricate optimization landscape, complicating the search for a global optimum. The variability in layer configurations $\mathbf{L}_e$ and $\mathbf{L}_d$ means that the effective architecture of the network changes with each configuration, making the parameter space significantly more complex.
Moreover, without proper regularization, the network overfits to specific layer configurations, reducing generalization and robustness. Dynamic activation further disrupts feature consistency, hindering shared representation learning. These challenges demand a training strategy that handles architectural variability while ensuring robust feature learning. 

\section{Solution Approach}\label{SA}
To effectively address the optimization challenges introduced by dynamic layer activation in the DD-JSCC architecture, we propose a solution that integrates Hierarchical Layer Activation and Implicit Regularization through sequential randomized layer selection during training. This approach simplifies the optimization landscape, prevents overfitting, ensures consistent feature representations, and stabilizes the training dynamics for both the encoder and decoder components of the network.
\vspace{-1mm}
\subsection{Hierarchical Layer Activation}
The dynamic activation of layers introduces an exponentially large number of possible layer configurations in both the encoder and decoder, given by $2^{L-3}$ for each (since the first two and the last layer are always active for the minimal configuration). To reduce this combinatorial complexity and ensure consistent feature representations, we introduce hierarchical constraints on the layer activations for both the encoder and decoder.
\subsubsection{Hierarchical Constraints}
For the encoder, we enforce that higher layers can only be active if all preceding layers are also active. Mathematically, the hierarchical constraint for the encoder is defined as:
\vspace{-2mm}
\begin{equation}
\vspace{-2mm}
l_i^e \leq l_{i-1}^e, \quad \forall i =  3, \dots, L-1,
\vspace{-2mm}
\label{eq:encoder_hierarchical_constraint}
\end{equation}
with
\vspace{-2mm}
\begin{equation}
l_1^e = l_2^e= l_L^e = 1,
\label{eq:encoder_critical_layers}
\end{equation}
where $l_i^e \in \{0,1\}$ indicates whether the $i$-th encoder layer is active ($1$) or inactive ($0$).
Similarly, for the decoder, we enforce:
\vspace{-2mm}
\begin{equation}
l_i^d \leq l_{i-1}^d, \quad \forall i =  3, \dots, L-1,
\label{eq:decoder_hierarchical_constraint}
\vspace{-2mm}
\end{equation}
with
\vspace{-2mm}
\begin{equation}
l_1^d = l_2^d = l_L^d = 1,
\label{eq:decoder_critical_layers}
\end{equation}
where $l_i^d \in \{0,1\}$ indicates whether the $i$-th decoder layer is active ($1$) or inactive ($0$). Under the hierarchical constraints, the number of possible layer configurations for both the encoder and decoder is reduced from exponential to linear.
\begin{algorithm}[!t]
\small
\caption{Training Procedure of the DD-JSCC Framework for Semantic Communication}
\label{alg:training_procedure}
\begin{algorithmic}[1]
\REQUIRE  Training dataset $\mathcal{D} = \{\mathbf{x}_i\}_{i=1}^N$, Maximum number of layers $L$, Learning rate $\eta$, Total epochs $E$, Batch size $B$, Power constraint $P_{\text{max}}$, SNR range $[\gamma_{\text{min}}, \gamma_{\text{max}}]$, CR range $[R_{\text{min}}, R_{\text{max}}]$.

\ENSURE 
Optimized encoder parameters $\phi^*$, Optimized decoder parameters $\theta^*$

\STATE Initialize encoder parameters $\phi$ and decoder parameters $\theta$ \quad \COMMENT{Only once}

\FOR{epoch $= 1$ to $E$}
    \FOR{each mini-batch $\{\mathbf{x}_b\}_{b=1}^B$ in $\mathcal{D}$}
        \STATE Randomly select $n \in \{3, \dots, L-1\}$  \quad \COMMENT{Random layer activation}
        \STATE Generate $\gamma$ uniformly from the range $[\gamma_{\text{min}}, \gamma_{\text{max}}]$
        \STATE Generate $R$ uniformly from the range $[R_{\text{min}}, R_{\text{max}}]$
        \vspace{-1mm}
        \STATE \textbf{Set Layer Configurations}:
        \[
        l_i^e = l_i^d = 
        \begin{cases}
        1, & \text{if } i \leq n \text{ or } i = L, \\
        0, & \text{otherwise}.
        \end{cases}
        \]
        \vspace{-2mm}
        \STATE \textbf{Encoder Forward Pass}:
        \STATE $\mathbf{h}^e = \mathbf{x}$ \quad \COMMENT{Input to the encoder}
        \FOR{$i = 1$ to $L$}
            \IF{$l_i^e = 1$}
                \STATE $\mathbf{h}^e = f_i^e(\mathbf{h}^e)$ \quad \COMMENT{Use stored weights for activated layers}
            \ENDIF
        \ENDFOR

        \STATE \textbf{Semantic Encoding}: $\mathbf{z} = \mathbf{h}^e$
        \STATE \textbf{Power Normalization}: $\mathbf{s} = \sqrt{\frac{K P_{\text{max}}}{\|\mathbf{z}\|^2}} \mathbf{z}$
        \STATE \textbf{Channel Transmission}: $\tilde{\mathbf{z}} = h \mathbf{s} + \mathbf{n}, \quad \mathbf{n} \sim \mathcal{CN}(\mathbf{0}, \sigma^2 \mathbf{I}_K)$

        \STATE \textbf{Decoder Forward Pass}:
        \STATE $\mathbf{h}^d = \tilde{\mathbf{z}}$ \quad \COMMENT{Decoder input}
        \FOR{$i = 1$ to $L$}
            \IF{$l_i^d = 1$}
                \STATE $\mathbf{h}^d = f_i^d(\mathbf{h}^d)$ \quad \COMMENT{Use stored weights for activated layers}
            \ENDIF
        \ENDFOR

        \STATE \textbf{Reconstruction}: $\hat{\mathbf{x}} = \mathbf{h}^d$
        \STATE \textbf{Loss Computation}: Using Eq. \ref{eq:loss_function}.
        \STATE \textbf{Update encoder parameters}: $\phi_{\mathbf{L}_e} \leftarrow \phi_{\mathbf{L}_e} - \eta \nabla_{\phi_{\mathbf{L}_e}} \mathcal{L}$
        \STATE \textbf{Update decoder parameters}: $\theta_{\mathbf{L}_d} \leftarrow \theta_{\mathbf{L}_d} - \eta \nabla_{\theta_{\mathbf{L}_d}} \mathcal{L}$
    \ENDFOR
\ENDFOR
\STATE \textbf{Return} Optimized parameters $\phi^*$, $\theta^*$
\end{algorithmic}
\end{algorithm}
This reduction applies separately to the encoder and decoder, but since we align their configurations during training (as will be discussed), the overall complexity is significantly reduced.

\subsubsection{Consistency in Feature Representations}
The hierarchical constraints ensure that each active layer in the encoder and decoder builds upon the outputs of the most recent preceding active layer, maintaining consistent feature representations across configurations. Instead of simply passing unchanged features through inactive layers, each active layer directly receives input from the last active layer in the sequence. Since the last layer is always active, if there is a gap due to skipped layers, the last active layer before this final layer directly provides its output to it.
For the encoder, the output of the $i$-th layer is given by:
\begin{equation}
\small
\vspace{-2mm}
\mathbf{h}_i^e = 
\begin{cases}
f_i^e(\mathbf{h}_{j}^e), & \text{if } l_i^e = 1, \quad j = \max \{k \mid k < i, l_k^e = 1\}, \\
\mathbf{h}_{j}^e, & \text{if } i = L \text{ and } \exists j < L \text{ such that } l_j^e = 1.
\end{cases}
\label{eq:encoder_layer_output}
\end{equation}
where $j$ represents the index of the most recent active layer before $i$, ensuring that inactive layers are effectively skipped. If no intermediate layers are active, the last active layer directly connects to the final layer. Similarly, for the decoder, the output of the $i$-th layer follows:
\begin{equation}
\small
\vspace{-2mm}
\mathbf{h}_i^d = 
\begin{cases}
f_i^d(\mathbf{h}_{j}^d), & \text{if } l_i^d = 1, \quad j = \max \{k \mid k < i, l_k^d = 1\}, \\
\mathbf{h}_{j}^d, & \text{if } i = L \text{ and } \exists j < L \text{ such that } l_j^d = 1.
\end{cases}
\label{eq:decoder_layer_output}
\end{equation}
\subsubsection{Alignment of Encoder and Decoder Configurations}
To ensure compatibility between the encoder and decoder, we align their layer configurations during training and inference:
\vspace{-3mm}
\begin{equation}
\mathbf{L}_d = \mathbf{L}_e.
\vspace{-3mm}
\label{eq:layer_configuration_alignment}
\end{equation}
This means that if a certain number of layers are active in the encoder, the same number of layers are active in the decoder, respecting the hierarchical constraints.

\subsubsection{Simplification of Optimization Problem}
By limiting the possible configurations and maintaining consistent computational graphs for both encoder and decoder, the hierarchical constraints simplify the optimization problem. They reduce the search space for the optimal parameters $(\phi, \theta)$ and mitigate issues related to training instability caused by varying network architectures.
\vspace{-3mm}
\subsection{Implicit Regularization through Sequential Randomized Layer Selection}
To prevent overfitting and improve generalization, we introduce an implicit regularization mechanism by employing sequential randomized layer selection during training for both the encoder and decoder.

\subsubsection{Expectation over Layer Configurations}
The optimization objective becomes an expectation over the randomly selected layer configurations $\mathbf{L}_e$ and $\mathbf{L}_d$ (which are aligned), leading to the following expected loss function:
\vspace{-2mm}
\begin{equation}
\small
\min_{\phi, \theta} \ \mathbb{E}_{\mathbf{x}, \gamma, R, \mathbf{L}_e, \mathbf{L}_d} \left[ \mathcal{L}\left( \mathbf{x}, D_{\theta} \left( h\mathbf{s} + \mathbf{n}; \gamma, R, \mathbf{L}_d\right) \right) \right],
\label{eq:expected_loss}
\vspace{-2mm}
\end{equation}
where $\mathbf{L}_d = \mathbf{L}_e$ due to the alignment.

\subsubsection{Implicit Regularization Effect}
The randomness in layer selection introduces variability during training for both the encoder and decoder, acting as an implicit regularizer. It discourages the model from relying too heavily on specific layers, thus reducing the risk of overfitting. This approach aligns with principles in statistical learning theory \cite{srivastava2014dropout}. The expectation over configurations introduces a regularization term in the loss function:
\vspace{-2mm}
\begin{equation}
\mathcal{L}_{\text{total}} = \mathbb{E}_{\mathbf{L}_e, \mathbf{L}_d}[\mathcal{L}] = \mathcal{L}_{\text{empirical}} + \lambda \Omega(\phi, \theta),
\vspace{-2mm}
\label{eq:regularized_loss}
\end{equation}
where \( \mathcal{L}_{\text{empirical}} \) represents the average reconstruction error over the training data, and \( \lambda \Omega(\phi, \theta) \) serves as a regularization term. This regularization arises from the variability introduced by random layer configurations during training. By averaging the loss over different configurations, the model is encouraged to learn parameters that perform consistently across various architectural setups, thereby preventing overfitting to specific configurations. 
Fig. \ref{system_model_overview} shows the architecture we used for the proposed mechanism and Algorithm 1 describes the whole solution approach for the proposed architecture.
\vspace{-2mm}
\subsection{Performance Metrics}
\vspace{-1mm}
To evaluate the effectiveness of the proposed DD-JSCC architecture, we employ the Peak Signal-to-Noise Ratio (PSNR) metric. It assess the quality of the reconstructed images compared to the original images.

\subsubsection{Peak Signal-to-Noise Ratio (PSNR)}
PSNR quantifies the reconstruction quality by measuring the ratio between the maximum possible pixel value and the Mean Squared Error (MSE) between the original image $\mathbf{x}$ and the reconstructed image $\hat{\mathbf{x}}$. A higher PSNR indicates better reconstruction quality.
\vspace{-2mm}
\begin{equation}
\text{PSNR} = 10 \cdot \log_{10}\left( \frac{{\text{MAX}_I^2}}{\text{MSE}} \right),
\vspace{-2mm}
\label{eq:psnr}
\end{equation}
where $\text{MAX}_I$ is the maximum possible pixel value (e.g., 255 for 8-bit images).
\begin{figure*}[t]
\centering
\includegraphics[width=18cm]{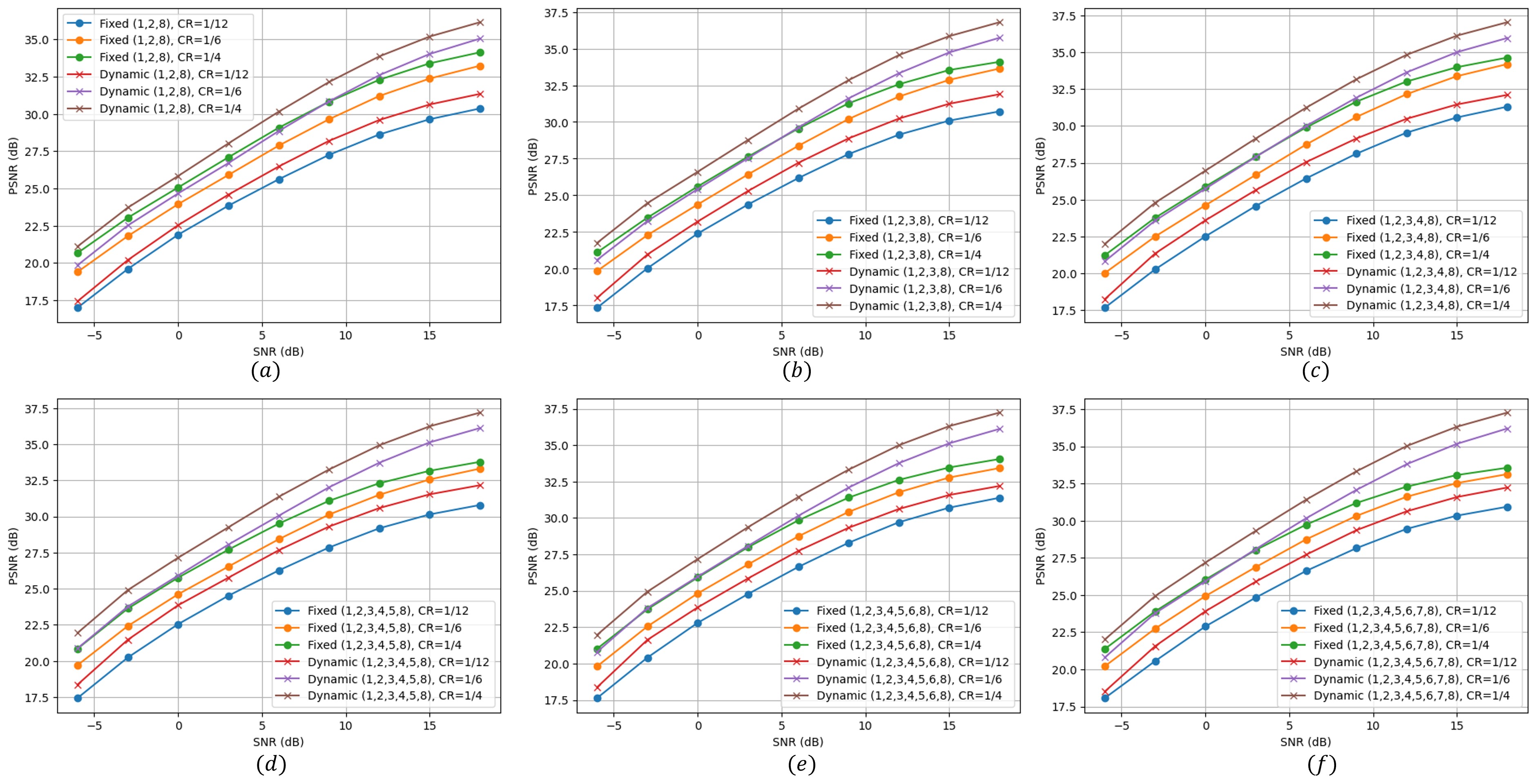}
\vspace{-2mm}
\caption{Performance Comparison Between the Proposed DD-JSCC and Fixed Deep-JSCC across Various Configurations.}
\vspace{-5mm}
\label{result01}
\end{figure*}
\begin{figure}[t]
\centering
\includegraphics[width=9cm]{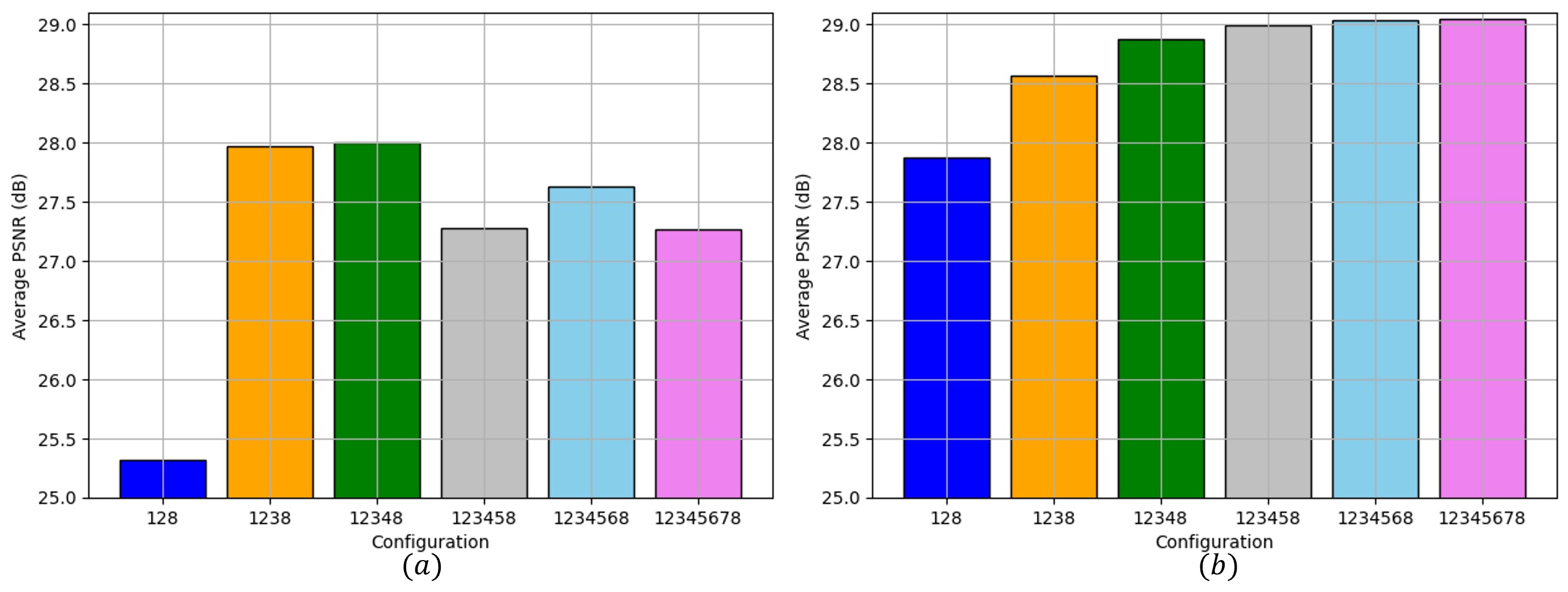}
\vspace{-5mm}
\caption{Average PSNR Comparison of DD-JSCC and Fixed Architectures Across Different Network Configurations.}
\vspace{-6mm}
\label{result_02}
\end{figure}

\section{Experimental Setup and Results}
We evaluated the proposed DD-JSCC against fixed architectures for wireless image transmission over an AWGN channel, using PSNR to assess reconstruction quality across different SNRs, CRs, and layer configurations. Each configuration (e.g., \texttt{Set12345678}, \texttt{Set12348}) denotes active layer subsets within the encoder-decoder. For instance, \texttt{Set12345678} activates all eight layers, while \texttt{Set128} uses only layers 1, 2, and 8.
\subsection{Dataset and Experimental Setup}
We implement the proposed models and benchmark schemes using PyTorch \cite{paszke2019pytorch}. To effectively capture visual differences among reconstructed images, we utilize the high-resolution DIV2K dataset \cite{agustsson2017ntire}, which is renowned in semantic communication community. The training subset of DIV2K is employed to train the network, while the testing subset is used to evaluate model performance. During training, models are trained across a range of SNR values, specifically $(0-27)$ dB. The channel encoders and decoders incorporate in total eight layers and are optimized using the Adam optimizer with a learning rate of $10^{-4}$. For compression, the model was trained with CRs ranging from 0.1 to 0.9, enabling it to handle a wide spectrum of compression levels. During testing, we evaluated the model at specific CRs of $\frac{1}{12}$, $\frac{1}{6}$, and $\frac{1}{4}$.
\subsection{Performance Analysis}
Figs. \ref{result01}(a) to \ref{result01}(f) illustrate the PSNR performance of both fixed and dynamic architectures across different sets of active layers, SNRs and CRs. Across all figures, the dynamic architecture consistently outperforms the fixed architectures. For instance, in fig. \ref{result01}(a), at a CR of $\frac{1}{12}$, the dynamic model achieved a PSNR of 31.35\,dB at 18\,dB SNR compared to the fixed model's 30.37\,dB, marking an improvement of approximately 0.98\,dB. Similarly, at an unseen SNR of $-6$\,dB and the same CR, the dynamic model attained a PSNR of 17.43\,dB versus the fixed model's 17\,dB, indicating a 0.43\,dB gain. These enhancements are even more pronounced at higher CRs; for example, at a CR of $\frac{1}{4}$ and 18\,dB SNR, the dynamic model reached 36.16\,dB compared to the fixed model's 34.14\,dB, resulting in a significant improvement of 2.02\,dB.

Furthermore, fig. \ref{result_02}(a) and fig. \ref{result_02}(b) present the average PSNR across all SNRs and CRs for different network configurations. The proposed DD-JSCC architecture demonstrates a consistent increase in average PSNR as the number of active layers increases, achieving an average PSNR of 29.04\,dB for the \texttt{Set12345678} configuration. In contrast, the fixed architectures exhibit irregular performance trends, with average PSNR values peaking at 27.99\,dB for the \texttt{Set12348} configuration and decreasing thereafter. This disparity highlights the effectiveness of our regularization and training techniques in DD-JSCC. The model shares parameters across all configurations and employs hierarchical layer activation with implicit regularization. As layers are added, shared representations and collective optimization prevent overfitting, while deeper layers refine features from shallower ones, enhancing pattern learning and performance.


The statistical analysis underscores the robustness and scalability of the DD-JSCC architecture. On average, the dynamic model achieved PSNR improvements ranging from 0.43\,dB to 2.02\,dB over fixed models across all configurations. These gains are particularly significant at higher CRs and SNRs, where maintaining image quality under stringent compression and challenging channel conditions is critical. Additionally, the DD-JSCC architecture's ability to generalize to unseen SNRs of $-6$\,dB and $-3$\,dB without performance degradation further validates the efficacy of the hierarchical layer activation and implicit regularization mechanisms employed during training.

Moreover, the training efficiency of the DD-JSCC architecture presents a practical advantage. While fixed architectures required separate training sessions for each layer configuration—accumulating to a total of 1,200 epochs for six configurations—the dynamic model achieved superior performance with just 700 epochs. This reduction in training time not only conserves computational resources but also facilitates quicker deployment and adaptability in real-world scenarios where channel conditions and compression demands may fluctuate unpredictably.

\vspace{-4mm}
\section{Conclusion}
\vspace{-2mm}
In this paper, we introduced DD-JSCC, a novel encoder-decoder architecture tailored for semantic communication systems. This is the first study to incorporate dynamic layer adjustments based on real-time assessments of transmitter and receiver capabilities, power constraints, compression ratios, and channel conditions. Our approach leverages hierarchical layer activation and implicit regularization through sequential randomized training, effectively reducing combinatorial complexity and preventing overfitting. This enables a single encoder-decoder framework to adaptively modify its structure, optimizing resource utilization and ensuring consistent feature representations across varying configurations. Simulation results demonstrate that DD-JSCC significantly enhances image reconstruction performance, achieving up to 2 dB improvement in PSNR over fixed Deep-JSCC architectures while reducing training costs by over 40\%. The unified and scalable framework of DD-JSCC eliminates the need for multiple specialized models, thereby reducing training complexity and deployment overhead.

\vspace{-4mm}
\bibliographystyle{IEEEtran}
\bibliography{references}
\vspace{-5mm}
\vspace{12pt}
\end{document}